# State-to-State Differential and Relative Integral Cross Sections for Rotationally Inelastic Scattering of H$_2$O by Hydrogen

C-H. Yang, G. Sarma, D. H. Parker, J. J ter Meulen

Radboud University Nijmegen, Institute for Molecules and Materials, Nijmegen, the Netherlands

L. Wiesenfeld

UJF-Grenoble 1 / CNRS-INSU, Institut de Planétologie et d'Astrophysique de Grenoble (IPAG) UMR 5274, Grenoble,

F-38041, France,

Abstract:

State-to-state differential cross sections (DCSs) for rotationally inelastic scattering of H$_2$O by H$_2$ have been measured at 71.2 meV (574 cm$^{-1}$) and 44.8 meV (361 cm$^{-1}$) collision energy using crossed molecular beams combined with velocity map imaging. A molecular beam containing variable compositions of the ($J$ = 0, 1, 2) rotational states of hydrogen collides with a molecular beam of argon seeded with water vapor that is cooled by supersonic expansion to its lowest para or ortho rotational levels ($J_{KaKc}$= 0$_{00}$ and 1$_{01}$, respectively). Angular speed distributions of fully specified rotationally excited final states are obtained using velocity map imaging. Relative integral cross sections are obtained by integrating the DCSs taken with the same experimental conditions. Experimental state-specific DCSs are compared with predictions from fully quantum scattering calculations on the most complete H$_2$O-H$_2$ potential energy surface. Comparison of relative total cross sections and state-specific DCSs show excellent agreement with theory in almost all details.

## 1. Introduction

In order to test the quality of potential energy surfaces (PES)s for $H_2O$-$H_2$ and $H_2O$-He at a collision energy relevant to astrophysical processes, we have recently reported rotationally resolved state-to-state differential cross sections for $H_2O$ collisions with $H_2$ and He using a crossed beam machine combined with velocity map imaging detection [1, 2]. For $H_2O$+He rotationally inelastic scattering [2], the state-to-state differential cross sections were extracted experimentally for the first time and were found to be in good agreement with full close-coupling quantum calculations based on a previously published [3] *ab initio* potential. In addition, a hard-shell ellipsoid model was employed to gain further physical insight in interpreting the observed rotational rainbows [4, 5, 6] observed in the $H_2O$-He differential cross sections. This article provides a full description of our studies on state-to-state differential cross sections of rotational excitation of $H_2O$ by the $H_2$ molecule.

Inelastic scattering probes the anisotropic part of the interaction potential, which is responsible for rotational energy transfer. [5, 7, 8, 9] Elastic [3, 10, 11, 12] and inelastic [1, 2, 3, 13] collisions with $H_2O$ have been studied in detail because of the general importance of water in many media including interstellar space. Elastic differential cross sections of $H_2O$ with the rare gases, $H_2$, and $H_2O$, were first acquired by Bickes *et al.* [11] The observed structures of diffraction oscillations and rainbow maximum for polar-non-polar interactions were used to determine the spherically symmetric model potential parameters (e.g. the well-depth and location of zero potential), assuming a Lennard-Jones (12,6) potential. Brudermann *et al* [3] presented measurements of differential cross sections for $H_2O$+He elastic scattering and reported partially state-resolved inelastic angular dependent energy loss spectra, at two different collision energies. Capelletti *et al.* [12, 14] reported determination of the potential parameters for the isotropic component of the $D_2O$-He and $D_2O$–$D_2$ interaction by elastic scattering studies. They compared their fitted parameters with those of previous studies using different potential models and with the predictions of *ab initio* calculated potential surfaces. Glory structures in the total cross sections for $H_2O$-rare gas collisions have also been investigated by the Perugia group.[15, 16] Rotationally resolved state-to-state integral cross sections of ortho and para-$H_2O$ collisions with Ar were investigated by Chapman et al. [13] They reported an approximate exponential-decay character of measured integral cross sections with the magnitude of rotational energy transfer of $H_2O$. In addition to the energy gap model, a clear propensity of rotational excitation of $H_2O$ around the *a* and *c* principal axes was observed and compared with classical and quantum calculations.

In this paper the collision processes

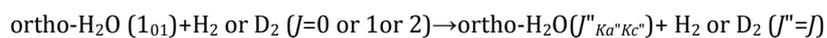

or

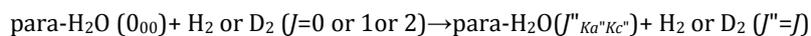

are studied. In our experiments, all collisional transitions occur in the ground vibrational states of $H_2O$ and hydrogen. Unprimed quantum states of $H_2O$ or hydrogen denotes levels before collision, and the double prime symbol " indicates nascent final states ($H_2O$ or hydrogen) after collision. A scheme of the $H_2O$ rotational energy level structure in its ground vibronic state ($v=0$, $\tilde{X}$) and possible rotational state transitions are presented in Fig. 1 which shows the rotational energy levels of ortho- and para-$H_2O$ separately, plotting the $Ka$ ladder. We use here the notation $J_{KaKc}$ for the rotational states of $H_2O$, with total angular momentum $J$ and the quantum numbers $Ka$ and $Kc$ are the values for the projection of $J$ on the $a$ and $c$ rotation axis. [17] In this computation, the rotational constants of $H_2O$ were taken at 27.88063134, 14.52176959, and 9.277708381 $cm^{-1}$. In addition, for the $H_2$ ($D_2$) molecule, the symbol $J$ is used to denote the rotational state. The rotational constant of $H_2$ is taken here at 60.853 $cm^{-1}$.

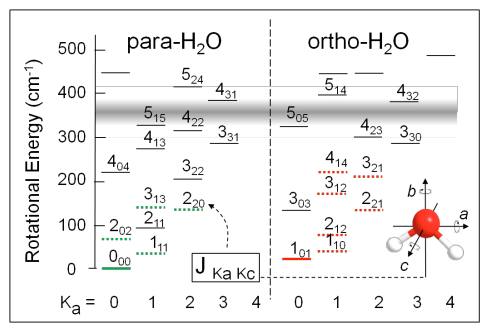

**FIG 1.** The rotational energy levels of ortho- and para-$H_2O$, plotted separately by the $Ka$ ladder. The shaded region shows the collision energy when using the $H_2$ beam with a 200K nozzle (collision energy 361 $cm^{-1}$). Ground state levels are indicated by a thick solid line while excited state levels populated by the collision and probed in this study are indicated by a thick dashed line. Principal rotational axes $a,b,c$ are labeled in the inset ball and stick model.

The presence of two identical H atoms in $H_2O$ results in two nuclear spin states: ortho ($K_a + K_c$ = odd), and para ($K_a + K_c$ = even). In addition to the conservation of total energy and momentum during the collisions, the rotational level transitions in $H_2O$ conserve nuclear spin, namely, ortho-to-para transitions for our collision conditions are forbidden.[13]  In addition, for the $H_2$ ($D_2$) molecule, with two identical hydrogen (deuterium) atoms, the coupling of nuclear spin results in ortho-$H_2$ ($D_2$) and para-$H_2$ ($D_2$). In electronic ground state of $H_2$ ($D_2$), states contain $J$ = even (odd) are para levels and $J$ = odd (even) are ortho levels. [18] The ortho/para ratios for thermodynamical equilibrium values at the high temperature limit are 3 for $H_2$ and 2 for $D_2$.

We use velocity map imaging [19] to measure state-to-state differential cross sections. The rotational states of $H_2O$ molecules before and after collisions were detected by a 2+1 Resonance Enhanced Multi-Photon Ionization (REMPI) process via the $\tilde{C}$-$\tilde{X}$ electronic transition. [17] Moreover, the rotational states ($J$) of hydrogen before and after inelastic scattering were examined by several different REMPI detection schemes. [20,21] The extracted DCSs from the experiments are compared with close-coupling calculations performed with state-of-the-art potential energy surfaces. [22]

The organization of this paper is as follows. Section 2 presents the details of experimental conditions, including the crossed beam machine, velocity map detector, ionization laser, REMPI spectroscopy, and preparation and detection of the cooled $H_2$ beam. Simple descriptions of image corrections, extraction of differential cross sections, and the theoretical calculations are presented in section 3. Section 4 presents experimental and calculated cross sections for different rotational transitions of $H_2O$ and $H_2$. The extracted and calculated DCSs for $H_2O$-hydrogen are compared and discussed in section 5. Section 6 contains summaries and concluding comments.

## 2. Experimental

The crossed beam experimental setup has been described briefly in our previous reports [1,2] and in more detail here. The setup, shown schematically in Fig. 2, can be presented in three parts: (1) two beam sources (primary and secondary beams) and their characterization by laser ionization, (2) the beam-crossing region, and (3) velocity map imaging detection:

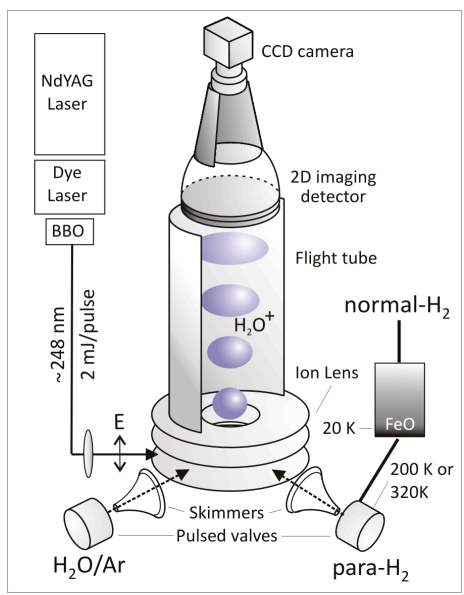

**FIG 2.** Schematic diagram of the crossed-beam velocity map imaging apparatus. A pulsed beam of $H_2O$ formed by seeding water vapor in Ar is skimmed and crossed by a second skimmed beam of pure normal- or para-$H_2$. The temperature of the secondary beam nozzle is fixed to either 200 or 320K. Rotationally excited $H_2O$ is state-selectively ionized by (2+1) REMPI via the $\tilde{C}$-state in the 248 nm region using a focused (20 cm lens) pulsed tunable dye laser beam that is frequency- doubled in a BBO crystal. The E field of the linear polarized laser beam lies perpendicular to the detector plane. The nascent $H_2O^+$ image is mass-selected by time-of-flight and projected onto a two dimensional (2-D) imaging detector then recorded by a CCD camera.

### 2.1. Beam Sources and collision properties

#### 2.1.a    *H₂O beam and laser ionization detection of the rotational state distribution*

The $H_2O$ (primary) beam was produced by flowing ~1 bar pure Ar gas into a simple $H_2O$ bubbler (demineralized water at 300K) and then through a hairpin-type pulsed valve (Jordan). The concentration of $H_2O$ in the primary beam was 2.5%, which was calculated assuming the vapor pressure of $H_2O$ at 300K. The $H_2O$ beam was rotationally cooled by adiabatic expansion and then collimated by a skimmer (Ø = 2.5 mm) located 30mm downstream from the valve exit. Analysis of the REMPI spectrum of the water beam before collision, as described next, indicates a rotational temperature of ~12 K, where more than 90% of ortho and 97% of para-$H_2O$ is populated in the lowest $1_{01}$ and $0_{00}$ rotational levels in the ground vibronic state, respectively. No experimental correction was made for the population of the $1_{10}$ state of ortho-$H_2O$ (~10%) or the $1_{11}$ state (~3%) in the primary beam. The effect of especially the $1_{11}$ state on the reported DCSs can be predicted by theory as shown in our previous paper on $H_2O$+He collisions.[2] Correction for these states was found here to be unnecessary for $H_2O$+$H_2$ collisions.

After collision, nascent $H_2O$ molecules were ionized by 2+1 REMPI via the $\tilde{C}$-$\tilde{X}$ transition as described in detail in a separate paper. [17] A tunable dye laser system (Lambda Physik ScanMate) pumped by the third harmonic of a pulsed Nd:YAG laser (Contiuum Powerlite 9010) was used for the ionization process. Coumarin 307 dye was used to produce dye laser wavelengths around 496 nm. Afterwards, the ionizing radiation (around 248 nm) was generated by frequency doubling the output of the dye laser with a BBO crystal. A typical power of the ionizing radiation was 2-3 mJ/pulse during the experiments. A 20 cm focal length spherical lens was used to focus the ionization laser beam to the center of the collision and ion optics region. The linear polarization of the ionization laser was kept perpendicular to the collision and detector planes as shown in Fig. 2. Changing the linear polarization of the laser from perpendicular to parallel to the collision plane did not cause any observably differences in the experimental results. Therefore, we ignore any alignment effects for our $H_2O$ product detection in our experimental conditions.

Previous studies have determined that linewidth broadening for the $\tilde{C}$-$\tilde{X}$ resonant REMPI transition is partially caused by heterogeneous predissociation of the $\tilde{C}$ $^1B_1$ Rydberg state, with the predissociation described by a simple $Ka'^2$ or $\langle Ja'^2 \rangle$-dependent model. [23], [24], [17] Here we use prime character, $'$ , to note the associated quantum numbers in the $\tilde{C}$ $^1B_1$ Rydberg. Consequently, the choice of REMPI transitions with a smaller value of $Ka'$ results in narrower REMPI lines, enabling the detection of single rotational states ($J"_{Ka"Kc"}$) of nascent $H_2O$ after collisions .

### 2.1.b. *Molecular hydrogen beam*

The $H_2$ (normal-$H_2$ and para-$H_2$) or normal-$D_2$ (secondary beam) was produced by a pulsed valve (Jordan) located in a separate source chamber with ~1bar stagnation pressure. After expansion the secondary beam was collimated by a skimmer (Ø = 3.0 mm) 30mm downstream from the valve. Normal $H_2$ (ortho:para=3:1), and normal $D_2$ (ortho:para=2:1) gases with 99.99% purity were commercially obtained and used without further purification.

Characterization of the rotational population of $H_2$ in its ground vibronic state has been investigated extensively in the past. [20, 21] In our experiments, the rotational state populations of $H_2$ were determined by either 3+1 REMPI via the $\tilde{C}$ ($v$=0)- $\tilde{X}$ ($v$=0) transition [21], or 2+1 REMPI via the E,F-X transition. [20] Four different hydrogen beam conditions were used; a supersonic expansion of normal $H_2$ using a 320 K nozzle and a nozzle cooled to 200K, and similar conditions for para-$H_2$. REMPI spectra were measured for each condition before and after collision with the $H_2O$ beam.   The REMPI measurements took place in the scattering center which confirmed that back conversion of para-$H_2$ to normal-$H_2$ did not take place in the pulsed valve. A typical (3+1) REMPI spectrum is shown in Fig. 3 for samples of normal-$H_2$ and our sample of para-$H_2$, using a 320 K valve.

The para-$H_2$ gas home-made conversion setup contained iron oxide powder at ~20K, cooled by liquid helium, which causes catalytic conversion of normal $H_2$ to para-$H_2$ ($J$=even). [25, 26]   Para-$H_2$ was stored in aluminum cylinders to slow down back conversion due to magnetic materials. The purity of para-$H_2$ was at least 90% (i.e. para-$H_2$:ortho-$H_2$ ~ 9:1). Because of the high separation of $H_2$ rotational states (rotational constant ~60 cm$^{-1}$), the adiabatic expansion of pure $H_2$ by the supersonic jet is not efficient in cooling $H_2$ ($J$=2) to $H_2$ ($J$=0).   In order to increase the amount of $J$=0 para-$H_2$, we used a modified liquid nitrogen cooled pulsed valve with an externally adjustable poppet (General valve) and a nozzle temperature cooled down to ~ 200K. The population of $J$=0 for this condition was estimated to be ~73% by the REMPI detection method. The rotational state populations for normal-$H_2$ and para-$H_2$ for the different nozzle temperatures are presented in Table 1.

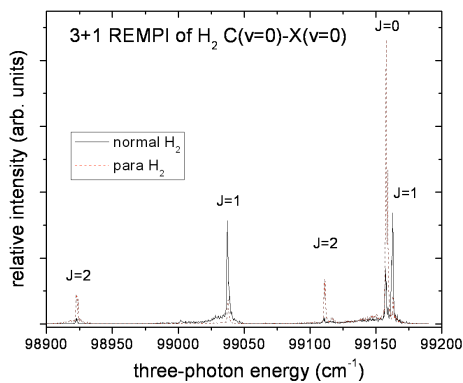

**FIG 3.** 3+1 REMPI spectrum of the ground vibronic state of $H_2$ via the $\bar{C}$ (v=0) $\leftarrow$ $\bar{X}$ (v=0) transition. The black line represents the spectrum when using normal-$H_2$ while the red line shows the spectrum of our prepared para-$H_2$. Both spectra were taken with a 320 K nozzle.

### 2.1.c. Collision conditions

Experimental conditions are presented in Table 2. The two molecular beams cross (90° angle) at the collision and ionization center, 90 mm downstream from both valves. With both beams on, the pressure in the collision chamber was ~3×10$^{-6}$ mbar. Measurements of the speed distributions of the molecular beams were performed by positioning the valves at different distances from the collision (ionization) center. By measuring the $H_2O$ signal of the molecular beams using Ar, He, or hydrogen carrier gases at the different valve-ionization distances, the speeds of different molecular beams can be approximately estimated from the valve-ionization distances and their corresponding arrival time delays. Residual $H_2O$ from the gas handling system of the secondary beam can disturb the scattering image. In order to eliminate residual water we tried several procedures. First of all, we pumped the whole gas system and cooled part of the gas line by liquid $N_2$ for at least 10 minutes before the experiment. Then, during the experiment operation, we still cooled the part of the gas line positioned before the valve by liquid $N_2$ and/or kept the collision partner gas flowing through the valve to an external exhaust. The speed of gas flow in the gas line to the exhaust was controlled by a gas pressure reducer and a needle valve.

The contribution of secondary collisions was evaluated by monitoring the intensity depletion of the $H_2O$ ground state signal. With temporal overlap of the two molecular beams, the conditions were adjusted (mainly via the backing pressure of the secondary molecular beam) such that less than 10% of $H_2O$ ground state was depleted by collisions. We can safely conclude that secondary collisions in our $H_2O$ inelastic scattering experiments can be ignored. Furthermore, the percentages of $H_2O$ clusters in the water beam was checked by scanning the mass gate time delay to higher m/e ranges than $H_2O^+$ while using up to 4 mJ/pulse of a focused laser beam (fl=20cm) at the wavelength on- or off-resonance with $H_2O$ REMPI lines. No significant water-water or water-Ar clusters in our molecular beams were observed at any laser wavelength in our tuning range. This is probably due to the rather warm operating conditions of the Jordan valve and the straight nozzle channel we used, which is not optimal for rotational cooling but which does disfavor cluster formation. In addition, the existence of collisions with clusters will cause different kinematics (e.g., via the reduced mass) and result in images with different locations and sizes of the Newton spheres. Non-monomer collisions can be evaluated by the image calibration methods.[27]    In our analyses, we found that the contribution of water-water or water-Ar clusters to our collision signals can be ignored.

Rotational state transitions in hydrogen molecules colliding with $H_2O$ (with the same restriction of ortho-to-ortho or para-to-para transitions) are in principle energetically allowed.   However, such $H_2$ state changing collisions were not observed experimentally.

### 2.2. Velocity map imaging detection

A set of electrode plates [19] (ion optics) was positioned in the center of the collision chamber with the ion optics and time-of-flight tube perpendicular to the collision plane. The ion optics were aligned carefully to set the focus point of the ionization laser and the collision center of the two beams at the center of the repeller and extractor plates, both vertically and horizontally. After collision and ionization, the $H_2O^+$ ions were extracted by the ion optics and pass through the time-of-flight tube. A typical setting for the repeller plate was 1000 Volt. The voltage ratio between the repeller and extractor plates were adjusted to obtain the best focus of the $H_2O^+$ molecular beam velocity on the MCP detector. [19] Afterwards, the Newton spheres of the state-selectively ionized $H_2O^+$ ions were crushed onto the MCPs, which are switched on at the appropriate moment. The whole collision experiment was operated in a pulsed mode at 10 Hz repetition rate. The time delays of the two molecular beams, ionization laser, MCPs, and camera were controlled by an 8-channel pulse/delay generator (BNC).

Signal from any remaining population of the detected final state in the $H_2O$ primary beam and from residual $H_2O$ in the collision chamber was eliminated using a background subtraction procedure. We changed the secondary valve time delay to control the temporal overlap with primary beam and recorded images under conditions with and without temporal overlap alternatively for every 1000~2000 laser shots. The temporal-separated images were subtracted from the temporal-overlap images to yield the final raw images. Here we assume that the small (at most a few percent) population of the detected final state in the primary beam does not contribute significantly to the measured scattering image. A typical accumulation time for one reliable image was 8000 laser shots at 10 Hz. The laser power and molecular beam conditions were continuously monitored during the experiments.

**Table 1.** Nozzle and rotational temperatures of $H_2$ molecular beam and the corresponding measured rotation populations. Uncertainty is ~5% in estimations of the rotational temperature of $H_2$.

| | Normal-$H_2$ | | Para-$H_2$ (90%) | |
|---|---|---|---|---|
| Nozzle Temperature | 320K | 200K | 320K | 200K |
| Rotational Temperature Of $H_2$ | 220($\pm$10)K | 170($\pm$10)K | 220($\pm$10)K | 170($\pm$10)K |
| $J$=0 | 17($\pm$0.5)% | 20($\pm$0.6)% | 61($\pm$2.2)% | 73($\pm$2.5)% |
| $J$=1 | 72($\pm$0.5)% | 74($\pm$0.3)% | 10($\pm$0.1)% | 10($\pm$0.1)% |
| $J$=2 | 8($\pm$0.5)% | 5($\pm$0.7)% | 29($\pm$2.1)% | 17($\pm$2.5)% |
| Higher $J$ | 3($\pm$0.5)% | 1($\pm$0.3)% | <1% | <1% |

**Table 2.** Molecular beam conditions and collision energy. Uncertainty is ~8% in speed ratio and velocity, and ~10% in collision energy.

| Nozzle conditions | Secondary beam | | | Primary beam |
|---|---|---|---|---|
| | He | $D_2$ | $H_2$ | $H_2O$ (seeded in Ar) |
| Nozzle pressure (bar) | 1 | 1 | 1 | 1 |
| Nozzle Temperature (K) | 320 | 320 | 320 200 | 320 |
| Peak velocity (m/s) | 1660 | 1660 | 2700 2100 | 625 |
| Speed ratio | 8 | 8 | 6 | 10 |
| Collision energy (cm$^{-1}$) | 430 | 430 | 574 361 | - |

## 3. Results

### 3.1 State-to-state DCSs for $H_2O$ collisions with normal- and para-$H_2$

A typical nascent $H_2O^+$ image for the $1_{01} \rightarrow 2_{12}$ transition is shown in Fig. 4, where the experimental geometry connecting the experimental lab and center of mass frames is also indicated. The asymmetry of the image with respect to the relative velocity is due to the need for density-to-flux transformation. A detailed description of image corrections and the extraction of differential cross sections has been presented in previous papers. [1, 2, 28, 29] In short, the flux-to-density correction was done using an apparatus-weighted function with iterative fitting or image division methods. In all cases the density-to-flux image correction program used to calculate the flux DCSs yielded a simulated density image in very good agreement (within 5% for all relevant image pixels) with the measured images. The extracted differential cross sections were normalized by setting the maximum intensity to unity, or by setting the integrated DCS intensity to the absolute integral (total) cross section from calculations.

The calculation of differential cross sections for $H_2O$-$H_2$ inelastic scattering has been presented in a previous publication.[1] We assume that both water and hydrogen molecules are rigid bodies, at their average geometries in their respective ground vibrational states. This approximation is valid for the collision energies studied here (< 600 cm $^{-1}$), since the first vibrational transition, the $H_2O$ bending mode is at 1595 cm$^{-1}$, resulting in a clear separation between the rotational and vibrational energies. The rigid-body, five-dimensional (5-D) potential energy surface (PES) $H_2O - H_2$ results from an averaging of the full 9-D PES that included internal water and hydrogen motion.[22] It has been recently shown[30] that averaging the full 9D PES over the molecular ground state wavefunctions or else taking this PES at the average values of the internal coordinates are two procedures totally equivalent at the precision we are looking for. The 5D PES is thus taken as the interaction of the two molecules, at their respective vibrational ground state average geometries.

All quantum scattering calculations were performed with the Molscat code at the Close Coupling level, using the diabatic modified log-derivative method of Manolopoulos.[31] The rotational basis set for water at the collisional energy of 570 cm$^{-1}$ is $J(H_2O) \leq 8$, $E_{rot} \leq 950$ cm$^{-1}$. Similarly for $H_2$ $J=0$, 2 (ortho) and $J=1$ (para) states were used in the channel basis. It has been repeatedly observed that inclusion of closed rotational channels is imperative in order to converge the closed-coupling calculations, with a special emphasis put on the $J=2$ level of $H_2$.[32, 33] The Molscat code delivers the $S$ matrix or $T$ matrix elements, which were subsequently combined to yield differential cross sections. The formulae were derived from ref. 34. Formula (1) gives the differential cross-section (DCS) as a function of the angle of deflection $\theta$, for an asymmetric top scattering off a rod:

$$\frac{d\sigma}{d\theta}\left(j''_1 \tau''_1 j''_2 \leftarrow j_1 \tau_1 j_2\right) = \frac{1}{4k^2[j_1][j_2]} \sum_{j'_1 j'_2, j_{12} j'_{12}} \sum_{j_{12}, l_1 l''_1 l'_2} P_\lambda (\cos\theta)(-1)^{\tau_{12} - j_{12}} \times$$
$$T^{*,l_1}\left(j''_1 \tau''_1 j''_2 j''_{12} l''_1; j_1 \tau_1 j_2 j_{12} l_1\right) T^{l'_2}\left(j''_1 \tau''_1 j''_2 j''_{12} l''_2; j_1 \tau_1 j_2 j_{12} l_2\right) Z\left(l_1 J_1 l_2 J_2; j_{12} \lambda\right) Z\left(l''_1 J_1 l''_2 J_2; j''_{12} \lambda\right)$$

(1)

In equation (1), $k$ is the collisional wave-number in the center of mass reference frame, $P_\lambda(\cos\theta)$ is the Legendre polynomial of order $\lambda$, $[j] = (2j+1)$, and the Z functions are Racah angular momentum coefficients, defined *e.g.* in Edmonds [35] equation (6.2.15). The quantum numbers go as follows. Double primed quantities denote values after collisions, unprimed ones before collision or else conserved quantities. The differential cross section is built by combining two different scattering amplitudes, with respective transition matrices $T''^{J_1}$ and $T''^{J_2}$, with * denoting complex conjugation. $J_1$ and $J_2$ are here the total angular momentum of the partial waves, the $l$ quantum numbers are the orbital quantum numbers. In this equation, angular momentum quantum numbers for water are $j_1\tau_1$, where $\tau$ is the proper second quantum number for a symmetric or asymmetric top. [36] Angular momentum of hydrogen is $j_2$. Both are coupled to form the angular momentum $j_{12}$.

The reduced collisional mass was taken as 1.812773730 a.m.u, All other Molscat parameters were taken at default values. The convergence criterion was of 1% for the inelastic cross sections, with a careful examination of the DCS convergence. Also, the variations of both total and differential cross sections were found to be very small (compared also to the experimental uncertainty) in the energy range $E_{coll} = 570 \pm 45$ cm$^{-1}$.

Five sets of differential cross sections of $H_2O$ inelastic collisions with respect to elastic or inelastic transitions of $H_2$ $J$ state including 0-0, 2-2, 0-2, and 2-0 transitions for para-$H_2$ and 1-1 for ortho-$H_2$ were calculated. In order to compare the experimental extracted DCSs with a mixture of angular momentum states for $H_2$, the calculated DCSs were averaged to the populations of $J$ of normal and para-$H_2$ with the weights presented in Table 1.

**FIG 4**. Schematic illustration of the experimental geometry for studying inelastic collisions of $H_2O$ with $H_2$ ($J$ = 0, 1, or 2). End points of the vectors of the $H_2$ beam and the relative velocity are out

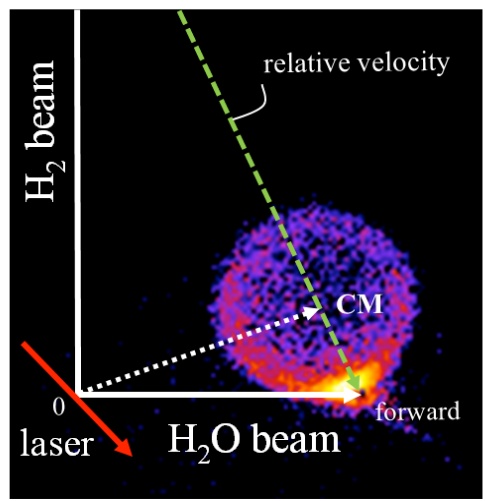

of the range of the figure. The presented $H_2O^+$ image is a 2-D projection of the Newton sphere formed by ortho-$H_2O$ collisions with para-$H_2$, for the $1_{01} \rightarrow 2_{12}$ transition. 'CM' stands for the position of the center-of-mass and '0' is the crossing point of the two molecular beams ($H_2O$ and $H_2$ beams) and the laser beam. The center-of-mass vector connects the center-of-mass and lab frames. Forward scattering is defined as scattered $H_2O$ molecules moving along the direction of the $H_2O$ CM velocity.

Our experimentally determined state-to-state DCSs for ortho- and para-$H_2O$ collisions with para-$H_2$ using a 320K temperature nozzle are presented together with the corresponding calculated differential cross sections in Fig. 5. All experimental DCSs show a very similar forward scattering angular distribution, and this feature is still dominant even for the highest rotational state of $H_2O$ that we could measure reliably, the $3_{30}$ state, in which more than 40% of the 574 cm$^{-1}$ collision kinetic energy was transferred to rotational energy. The CM angular distributions for almost all transitions also show a quite smooth decrease over the $60^0$ to $180^0$ deflection angles where rotational rainbows usually appear, especially for the higher final rotational states. The fluctuations in the shape of the differential cross section at this range of deflection angles for some of the weaker final states (e.g. the $3_{12}$, $4_{14}$, and $3_{13}$ states) are attributed to the low signal to noise ratio in the images.

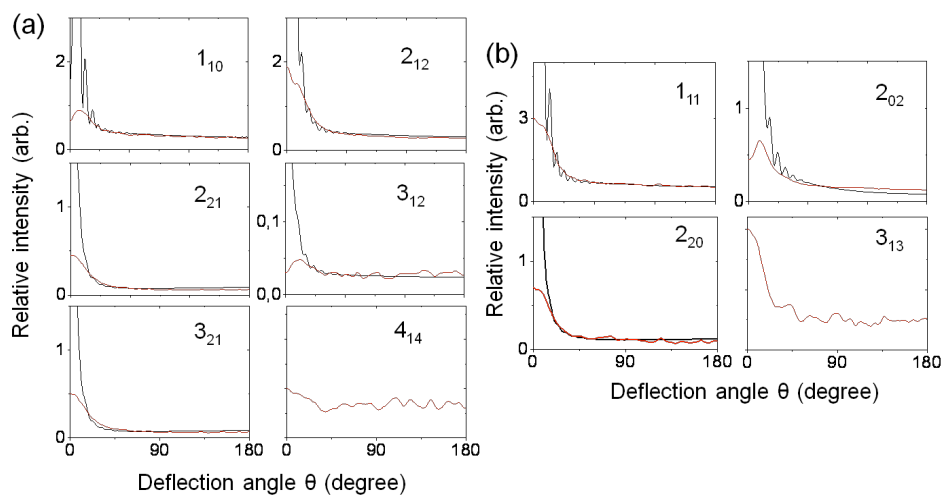

**FIG 5.** Experimentally extracted (red curve) and calculated (black curve) state-to-state differential cross sections for $H_2O$ collisions with para-$H_2$ expanded in a 320K nozzle (collision energy 574 cm$^{-1}$). (a) ortho-$H_2O$ (b) para-$H_2O$. The experimental curves were scaled to match the theoretical curve at a deflection angle of 60°. Experimental uncertainties are estimated as <10% at all angles >5° for the stronger transitions and <20% for the weaker final states, e.g. the $3_{12}$, $4_{14}$, and $3_{13}$ states.

Figure 5 shows results for scattering of water with para-$H_2$ using a 320K nozzle. For the other experimental conditions (para-$H_2$ with a 200K nozzle, normal-$H_2$ with a 320K and 200K nozzle) although the $H_2$ $J$ state populations changed dramatically, the experimental DCSs retain their main feature, namely dominant forward scattering. Experimental DCSs were scaled to theory at the scattering angle of 60 degrees, which was chosen as a region in the image away from the forward scattering peak yet still a region where the signal strength was good. The forward scattering region is less reliable experimentally due to uncertainties resulting from background subtraction, especially for the lowest energy final states where a small population is still present in the parent $H_2O$ beam. Agreement between theory and experiment is very good in all regions away from forward scattering (30-180°). Less satisfactory agreement occurs in the very forward angular range, where the experiment is less reliable, but where theory appears to consistently overestimate the amount of scattering.

### 3.2 State-to-state relative integral cross sections

Relative state-to-state integral cross sections for $H_2O$-$H_2$ inelastic collisions have been extracted from our corresponding experimental state-to-state DCSs (each taken under the same collision conditions) by first integrating the experimental images over all deflection angles to obtain the state-to-state relative total signals, which were then corrected by the line strengths for 2+1 REMPI detection of $H_2O$. Corrections of the differential cross sections for their different recoil velocity in the lab frame is not necessary in the case of $H_2O$-$H_2$ collisions because of their small Newton spheres and similarly shaped DCSs for all transitions.[7] Experimental state-to-state relative integral cross sections for ortho- and para-$H_2O$ colliding with para-$H_2$ using a 320K nozzle are presented in a bar graph in Fig. 6, where the relative integral cross sections, as expected, decrease qualitatively with the increasing degree of energy transfer.    The energy spacing from the ground state to the indicated final state is shown in Fig. 1 and also listed in Table 3.    A clear exception to the exponential fall-off in energy is the $2_{11}$ final state, which is discussed later.

Calculated state-to-state integral inelastic cross sections, presented in Table 3, were obtained from the calculated DCSs by integrating the differential cross sections, weighted by sin θ, over all deflection angles. Four selected transitions for each initial rotational state ($1_{01}$ and $0_{00}$ of ortho- and para-$H_2O$, respectively) are presented in Table 3. For $H_2$ rotational state transfer two para ($J$=0 and 2) and one ortho ($J$=1) states are considered and consequently five possible calculated transitions are presented. The trends in cross sections are relative, not only with regard to the magnitude of $\Delta J$, $Ka$ and $Kc$, and the energy transfer of $H_2O$ but also with the initial and final $J$ and $\Delta J$ for $H_2$. In general, the integral cross section decreases when the magnitude of energy transfer increases for both ortho- and para-$H_2O$.

Comparisons of experimental and calculated relative integral cross sections are presented in Fig. 6 by normalizing the experimental relative cross sections to theory at the $1_{11}$ final state. The calculated relative integral cross sections for the comparisons were obtained by averaging cross sections using the relative ratios of $H_2$ $J$ state populations presented in Table 1. The experimental uncertainties of the relative image intensities measured on different days result in 20% error bars for the integral cross section results.

The measured relative integral cross sections agree within the error bars in almost all cases with the quantum mechanical calculations.

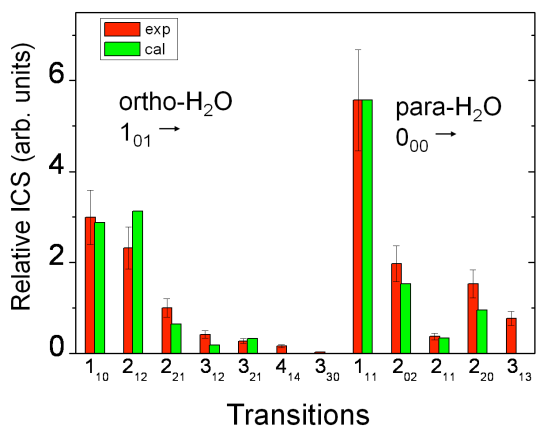

**FIG 6**. Comparison of experimental relative integral cross sections with quantum mechanical calculations for collisions of ortho- and para-$H_2O$ with para-$H_2$ (320K nozzle). The experimental relative cross sections are normalized to the calculated cross sections at the $1_{11}$ state. The experimental uncertainty is ∼ 20% .

**Table 3:** Calculated integral state-to-state cross section values for $H_2O$ + $H_2$ inelastic transitions, at kinetic energy $E_{coll}$ = 575 cm$^{-1}$ (320 K $H_2$ nozzle). For each $H_2O$ state-to-state transition the cross section (in Å$^2$) for each $H_2$ state-to-state component is listed separately. The total $H_2O$ state-to-state cross section shown in Figure 6 are calculated for each $H_2O$ component by summing over the separate $H_2$ cross sections weighted by the $H_2$ state populations listed in Table 1.

| $H_2O$ transition $J_{Ka\,Kc} \rightarrow J''_{Ka''\,Kc''}$ Energy level (cm$^{-1}$) | $H_2$ transition $J \rightarrow J''$ | Cross section (Å$^2$) | $H_2O$ transition $J_{Ka\,Kc} \rightarrow J''_{Ka''\,Kc''}$ Energy level (cm$^{-1}$) | $H_2$ transition $J \rightarrow J''$ | Cross section (Å$^2$) |
|---|---|---|---|---|---|
| *ortho* $\mathbf{1_{01} \rightarrow 1_{10}}$ 23.799 42.402 | $0 \rightarrow 0$ | 4.22 | *Para* $\mathbf{0_{00} \rightarrow 1_{11}}$ 0 37.158 | $0 \rightarrow 0$ | 8.09 |
| | $2 \rightarrow 2$ | 7.68 | | $2 \rightarrow 2$ | 15.10 |
| | $1 \rightarrow 1$ | 8.26 | | $1 \rightarrow 1$ | 15.71 |
| | $0 \rightarrow 2$ | 0.074 | | $0 \rightarrow 2$ | 0.061 |
| | $2 \rightarrow 0$ | 0.167 | | $2 \rightarrow 0$ | 0.307 |
| $\rightarrow \mathbf{2_{12}}$ 79.513 | $0 \rightarrow 0$ | 4.49 | $\rightarrow \mathbf{2_{02}}$ 70.133 | $0 \rightarrow 0$ | 1.39 |
| | $2 \rightarrow 2$ | 8.48 | | $2 \rightarrow 2$ | 5.42 |
| | $1 \rightarrow 1$ | 9.12 | | $1 \rightarrow 1$ | 5.70 |
| | $0 \rightarrow 2$ | 0.062 | | $0 \rightarrow 2$ | 0 |
| | $2 \rightarrow 0$ | 0.269 | | $2 \rightarrow 0$ | 0.631 |
| $\rightarrow \mathbf{2_{21}}$ 135.322 | $0 \rightarrow 0$ | 0.562 | $\rightarrow \mathbf{2_{11}}$ 95.245 | $0 \rightarrow 0$ | 0.018 |
| | $2 \rightarrow 2$ | 2.22 | | $2 \rightarrow 2$ | 1.67 |
| | $1 \rightarrow 1$ | 2.71 | | $1 \rightarrow 1$ | 1.51 |
| | $0 \rightarrow 2$ | 0.011 | | $0 \rightarrow 2$ | 0 |
| | $2 \rightarrow 0$ | 0.199 | | $2 \rightarrow 0$ | 0.202 |
| $\rightarrow \mathbf{3_{21}}$ 212.628 | $0 \rightarrow 0$ | 0.265 | $\rightarrow \mathbf{2_{20}}$ 136.588 | $0 \rightarrow 0$ | 0.784 |
| | $2 \rightarrow 2$ | 1.19 | | $2 \rightarrow 2$ | 3.34 |
| | $1 \rightarrow 1$ | 1.36 | | $1 \rightarrow 1$ | 4.09 |
| | $0 \rightarrow 2$ | 0.002 | | $0 \rightarrow 2$ | 0.006 |
| | $2 \rightarrow 0$ | 0.501 | | $2 \rightarrow 0$ | 0.509 |

### 3.3 Inelastic scattering of $H_2O$ by $D_2$

State-to-state differential cross section for $H_2O$ collisions with normal-$D_2$ have been investigated for several purposes. First of all, normal-$D_2$ has a quite different $J$ state population distribution than normal-$H_2$ due to its different nuclear spin statistics. The ratio of ortho ($J$=even) to para ($J$=odd) of normal $D_2$ is 2:1 at 320K and should remain the same after supersonic expansion. Consequently, $J$=0 dominates the initial $J$ state population of $D_2$ in the collisions with $H_2O$. Secondly, according to our previous publications, [1,2] the DCSs of $H_2O$-He are dramatically different from those shown here for $H_2O$-$H_2$ inelastic collisions. It is interesting to check that the difference in collision dynamics is not just a mass effect. $D_2$ has the same mass as He, but clearly the PES for $H_2O$+normal-$D_2$ is quite different from that of $H_2O$-He. [22, 37, 38] Experimental DCSs for $H_2O$+normal-$D_2$ are presented in Fig. 7. Here we mainly investigate the lower rotational state transitions for ortho- and para-$H_2O$. Obviously, $H_2O$ + normal-$D_2$ inelastic collisions show a similar behavior to $H_2O$-$H_2$ collisions, in which forward scattering again dominates the angular distribution of scattered $H_2O$ molecules.

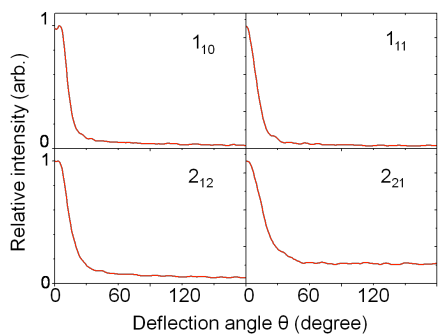

**FIG 7**. Experimental state-to-state differential cross sections of ortho- and para-$H_2O$ colliding with normal-$D_2$, collision energy 430 cm$^{-1}$. All signals were normalized to set each maximum intensity to unity.

# 4. Discussion

## 4.1 Differential cross sections for $H_2O$ collisions with para- and normal-$H_2$

Because full quantum theory has been shown here to be successful in predicting the main features of the experimental DCSs, we look now into the $H_2$ $J$-state dependences of the DCSs predicted by theory in order to understand the observed preference for forward scattering. Calculation results presented in Table 3 show that $H_2O+H_2$ collisions are mainly elastic with respect to $H_2$ for both para-$H_2$ ($J$=0, 2) and ortho-$H_2$ ($J$=1) scattering. In other words, $J$-state conserving instead of $J$-state changing $H_2$ rotation is the main process in $H_2O$-$H_2$ collisions. A similar conclusion was obtained in our experimental measurements monitoring the $J$-state populations of $H_2$ after collision using REMPI detection of $H_2$. The importance of $H_2$ $J$-state conservation during $H_2O$-$H_2$ inelastic collisions is attributed to the large rotational state spacing of $H_2$ compared to the collision energy and the state spacing in $H_2O$. Therefore, in our description of the shape of the DCS we use only theoretical DCSs for three elastic collisions of the $H_2$ molecule, namely $0{\rightarrow}0$, $2{\rightarrow}2$, and $1{\rightarrow}1$, as presented in Table 3 for different experimental conditions (i.e. para- and normal-$H_2$ using different temperatures for the pulsed valve nozzle). The shapes, relative ratios and intensities of the DCSs with respect to these three elastic processes for $H_2$ will determine the structures in the angular distributions for $H_2O+H_2$ scattering.

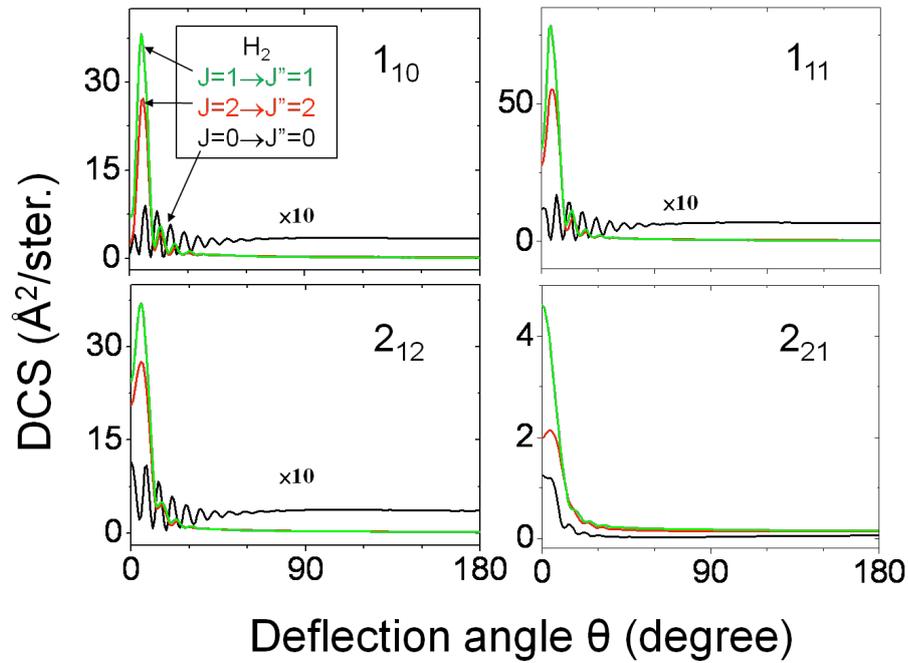

**FIG 8.** Theoretical DCSs in Å$^2$/steradian for $H_2O+H_2$ scattering where four final rotational states for ortho- ($1_{01} \rightarrow 1_{10}, 2_{12}, 2_{21}$) and para- ($0_{00} \rightarrow 1_{11}$) $H_2O$ scattering are plotted for three different elastic events for $H_2$, namely $J=0 \rightarrow J''=0$ (black curve), $2 \rightarrow 2$ (red curve), and $1 \rightarrow 1$ (green curve). The $J=0 \rightarrow J''=0$ (black) curve is multiplied by 10 for the $1_{10}$, $1_{11}$, and $2_{12}$ final states for clarity.

Theoretical DCSs for the three elastic processes involving $H_2$ are presented in Fig. 8. DCSs for the two $J>0$ processes show similar structure and intensity while $J=0$ presents a very different angular distribution (except for $2_{21}$) and much weaker intensity, in line with all previous calculations. The angular distribution for the $H_2$ $J>0$ processes peaks extremely in the forward direction (i.e. small deflection angles), while in the $H_2$ $J=0$ process for the three lower state transitions of ortho- and para-$H_2O$ presented in Fig. 8, i.e. $1_{01} \rightarrow 1_{10}$, $1_{01} \rightarrow 2_{12}$, and $0_{00} \rightarrow 1_{11}$, in addition to forward scattering, the sideways and backward parts contribute significantly to the angular distributions. Note, however, the strong forward scattering for the $2_{21}$ state predicted by theory, even for the $H_2$ $J=0 \rightarrow J''=0$ component.

According to Table 1, using a 320K nozzle, the para-$H_2$ molecular beam contains about 61% $J=0$ while a normal-$H_2$ beam has ~17% $J=0$. Consequently, for our experimental conditions, the DCSs for para-$H_2$ collisions with $H_2O$ should present higher contributions to the sideways and backward angular distribution than the DCSs for a normal-$H_2$ collision. In Fig. 9, theory is compared with experiment using the 320K value, where the theory curves are a summation of the curves shown in Fig. 8, weighted by the $J$-state populations. In Fig. 9, we notice that for each $H_2O$ rotation state transition the contributions of the side-backward angular distribution in the DCSs for para-$H_2$ are quite different from those of normal-$H_2$ for both the experimental observations and theoretical calculations, due to different contributions (ratios) from $H_2$ $J=0$.

The calculated and experimental DCSs of $H_2O$-$H_2$ inelastic collisions are seen from Figures 8 and 9 to show a strong forward scattering preference for all rotation state transitions of $H_2O$ due to the domination of the $H_2$ $J>0$ processes. The $H_2$ $J>0$ states possess a quadrupole moment, resulting in a dipole-quadrupole $H_2O$-$H_2$ interaction, which should be the main contribution to the observed forward scattering. Calculation, however, overestimates the scattering intensity for small deflection angles compared with the measurements, which might indicate an inaccuracy of the PES at large impact parameters.

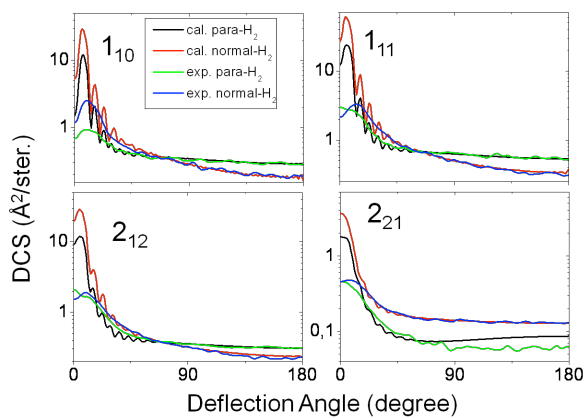

**FIG 9**. Comparison of calculated and experimental DCSs for para- and normal-$H_2$ collisions with ortho- and para-$H_2O$ using the pulsed $H_2$ valve at 320K. Calculated DCSs: para-$H_2$ (black curve) and normal-$H_2$ (red curve); experimental DCSs: para-$H_2$ (green curve) and normal-$H_2$ (blue curve). Theory curves are in absolute units, while the experimental curves were scaled to match the theoretical curve at the deflection angle of 60°.

## 4.2 Features of the relative integral cross sections for $H_2O$-$H_2$ collisions

Figure 6 shows qualitatively that the state-to-state relative integral cross sections decrease with increasing rotational excitation, in accord with the energy gap rule.[13]   A least-squares fit of the cross section versus the amount of rotational energy transfer $\Delta E$ for ortho-$H_2O$ + para-$H_2$ using a 320K nozzle to the function $\sigma(\Delta E)= \sigma_0 \exp(-\Delta E/E_0)$ yields a value of $E_0=78(\pm9)$ cm$^{-1}$. The extracted exponential decay rate from our $H_2O$-$H_2$ measurements is quite similar to that measured for $H_2O$-Ar scattering in a previous publication ($E_0=82$ cm$^{-1}$).[13] A small difference in $E_0$ for the two systems may indicate less shielding effects for collisions with $H_2O$ for the smaller size and lower mass of $H_2$ compared with Ar. [39]

Cross sections also depend on the rotational axes of $H_2O$. Excitation of rotation around these axes, which are labeled by the *Ka* and *Kc* quantum numbers in Figure 1, may follow propensity rules, as can be explained by two different approaches. The first on is based on the intramolecular body-fixed alignment effect.[13, 39] Good examples to describe this effect are the total cross sections to the $2_{02}$, $2_{11}$, and $2_{20}$ states of para-$H_2O$, plotted in Fig. 6. The $2_{02}$, $2_{11}$, and $2_{20}$ states represent rotation of $H_2O$ around the *c*, *b*, and *a* axes, respectively. In our measurements, we found that the relative cross section for the transition to the $2_{11}$ state (*b* axis) is much smaller than the one for the other two transitions. Following the discussion based on a classical mechanics model of Kolb *et al*.[39] impact on the O atom of $H_2O$ by the collision partner (here $H_2$) cannot rotate the $H_2O$ molecule around the *b* axis because (1) the perpendicular line to the O atom surface at the impact point always passes through the center of the O atom and (2) the center of the O atom and the center of mass of $H_2O$ are both on the *b* axis. Consequently, only impact on the H atoms of $H_2O$ may cause rotation around the *b* axis. It should be added that the O atom is so much larger than the H atoms that collisions with the O atom are much more probable than collisions with the H atoms. This means that the probability of $H_2O$ rotation around the *b* axis is smaller than that around the *a* and *c* axes. This preference of $H_2O$ rotation around the A or C axis after collision over rotation around the *b* axis has also been observed for $H_2O$-Ar collisions.[13]

We can also discuss propensity rules by examining the dipole allowed transitions. The $H_2O$ – $H_2$ PES main anisotropic terms all stem from the large dipole moment of water, 1.85 Debye, aligned with the symmetry *b*-axis. The dipolar transition rules are as follows, for $H_2O$ : $\Delta J = 0; \pm 1$ and $\Delta Ka = \pm 1; \pm 3$, $\Delta Kc = \pm 1; \pm 3$. This is clearly seen in Table 3 and Fig. 6, where the $\Delta J > 1$ transitions are clearly disfavored. If $\Delta J = +1$, the dipole rules apply and the $1_{01} \rightarrow 2_{12}$ transition in favored with respect to $1_{01} \rightarrow 2_{21}$, by nearly one order of magnitude. However, for $\Delta J = +2$, the transition is no more dipolar, and the situation is no longer clear-cut. If at a first approximation the transition is of a quadrupole type, then $\Delta J = 0, \pm 2$, and the wave functions keep their symmetries, with $\Delta Ka = \pm 2$, $\Delta Kc = \pm 2$. This propensity is observed theoretically and experimentally. Induced dipole and dispersion interactions with different selection rules, however, could also play a role.

One noticeable result from the calculations is that the $2 \rightarrow 0$ transition of the $H_2$ *J* state becomes more important when the amount of rotational energy transfer in $H_2O$ is larger (e.g., $1_{01} \rightarrow 3_{21}$ transition). This process should be more easily observed in the DCSs than in the relative integral cross sections because of the relatively small population of the $J=2$ initial state of $H_2$ and the relatively higher cross sections of the *J* conserving mechanisms. However, a larger ring on the imaging detector, indicating a larger amount of energy transfer, is not evident in the $H_2O$ image, perhaps because (1) the (calculated) shape of the DCSs for $H_2$ $2 \rightarrow 0$ is quite similar to those of the $H_2$ $J>0$ elastic processes and (2) the different radius of the Newton sphere due to the $H_2$ $2 \rightarrow 0$ process cannot be distinguished due to the limited energy resolution for these small images.

Due to the large difference in the integral cross sections between $H_2$ $J=0$ and $J>0$ elastic processes as presented in Table 3, the relative integral cross sections of $H_2O$ colliding with para-$H_2$ should be smaller than those for normal-$H_2$, which has a much higher initial population of $H_2$ $J=1$. We performed experimental measurements for determining the ratios of the state-to-state relative cross sections for collisions with normal-$H_2$ versus para-$H_2$. Experimental ratios (with 20% error bar) together with calculations for two different nozzle temperatures are presented in Fig. 10. Experiment and theory agree quite well at least for three lower rotation states of $H_2O$, even though the actual collision energy of the 200K nozzle experiment (361 cm$^{-1}$) is smaller than that used for the calculations (574 cm$^{-1}$).

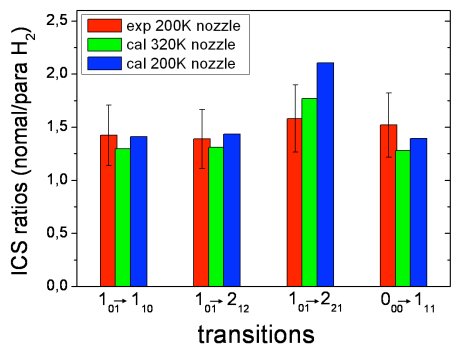

**FIG 10.** Experimental and calculated ratios of state-to-state relative cross section of $H_2O$ collision with normal- and para-$H_2$ using a 200K and 320K nozzle. The experimental results have 20% error bars and ~361 cm$^{-1}$ collision energy. The calculations consider rotational population for two different nozzle temperatures and collision energy of 575 cm$^{-1}$.

### 4.3 Comparison of the $H_2O$-He and $H_2O$-hydrogen collision systems

$H_2$ and He are two of the most abundant particles involved in collisions with $H_2O$ in interstellar space. State-to-state differential cross sections for $H_2O$-He and $H_2O$-$H_2$ inelastic collisions are quite different, as noted in our previous publication. [1] As discussed in Ref. 1, results for the relatively simpler $H_2O$-He system have been used previously, for estimating collision rates and cross sections for the more complicated $H_2O$-$H_2$ system. The large differences found between $H_2O$-He and $H_2O$-$H_2$ DCSs warrants extra caution in relating the two systems. In this work we also performed $H_2O$-$D_2$ collisions with normal-$D_2$, which has the same mass as He and a higher (but not exclusive) population of the $J$=0 ground rotation state than that of normal-$H_2$. DCS results for $H_2O$-$D_2$ still show a very similar forward scattering preference, like that of $H_2O$-$H_2$. We conclude that the difference in the DCSs for $H_2O$-He versus $H_2O$-$H_2$ arises not from a mass effect but from the dominant participation of $H_2$ $J$>0 processes, based on the predictions of the theoretical calculations.

From the above discussion we should expect that collisions of water with $H_2$ limited to the $J$=0 state will show similarities with water-He collisions. Figure 11 presents comparisons of calculated state-to-state integral cross sections for $H_2O$ inelastic collisions with $H_2$ ($J$, 0-0) and with He, at different collision energies. The calculations of $H_2O$+$H_2$ is from this work using the $J$=0,2 basis set but considering only the 0→0 process for the $H_2$ $J$ state, at a collision energy of 574 cm$^{-1}$. We also show results of previous calculations for $H_2O$+He from Patkowski $et$ $al$ [37] using two collision energies (464 and 566 cm$^{-1}$). From this comparison, the state-to-state integral cross sections for the two systems are indeed quite comparable. Basic similarities of the shapes, but with different phases of oscillations are also found between the differential cross sections for water-$H_2$($J$=0) and water-He.

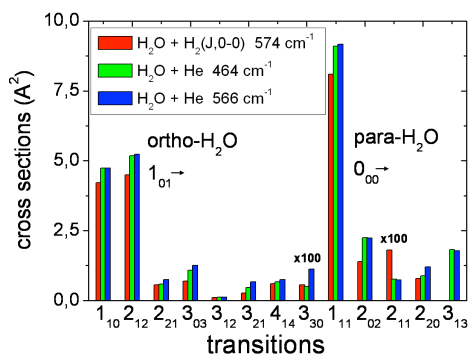

**FIG 11.** Comparisons of calculated integral cross sections for H$_2$O inelastic collisions with H$_2$ ($J$, 0-0) and He at different collision energies. The calculation of H$_2$O+H$_2$ is from this work while calculations for H$_2$O+He are from Patkowski *et al.*[37]

The strong similarity in integral (and differential) cross sections water-H$_2$($J$=0) and water-He is at first sight surprising for several reasons. First of all, the PES of H$_2$O-H$_2$ is quite different from that for H$_2$O-He. The PES of H$_2$O-H$_2$ has two minima which are located at 0$^{\cdot}$ (R=5.82 Bohr radius, E=-235.14 cm$^{-1}$ for the 5-D PES) and 119$^0$ (R=6.07 Bohr radius, E= -199.40 cm$^{-1}$ for the 5-D PES) with respect to C$_2$ axis of H$_2$O in the H$_2$O plane. [22] The PES of H$_2$O-He for our comparisons has only one minimum (R=5.92 Bohr radius, E= -34.9 cm$^{-1}$) which is located at 75$^{\cdot}$ with respect to the H$_2$O C$_2$ axis (O atom side) in the H$_2$O plane. In addition, the ratio of the collision energy to the energy depth of the potential well is much smaller for H$_2$O-H$_2$ experiments (2.45) than that for H$_2$O-He experiments (12.29). For H$_2$O-He, mainly the repulsive wall of the PES determines the collision dynamics due to this large (collision energy/potential well) ratio. Apparently, once this ratio exceeds unity the two systems probe similar parts of the potential energy surface. The importance of the multipolar, and particularly the dipolar, terms tend to show that the long distance anisotropic part of the intermolecular potential dominates, but further detailed theory  is needed to strengthen this assertion.

## 5. Conclusions

Rotationally resolved state-to-state differential and relative integral cross sections for $H_2O$-hydrogen inelastic collisions at 574 cm$^{-1}$ collision energy in the center of mass system have been investigated in detail using a crossed beam machine with velocity map imaging detection. The experimental cross sections agree extremely well with quantum calculations using the most complete $H_2O$-$H_2$ PES of Valiron *et al.* [22] For the differential cross section results we found a strong preference for forward scattering in the angular distributions for all measurable $J''_{Ka''Kc''}$ final states for collision with $H_2$ and $D_2$ at different experimental conditions. This forward scattering preference is attributed to a dominant participation by $H_2$ $J>0$ with $J = J''$ ($J$ state conserving), from the prediction of calculations. Side and backward contributions to the DCSs came only from $H_2O$-$H_2$($J = J''=0$), as has been confirmed by comparisons of $H_2O$+normal-$H_2$ and $H_2O$+para-$H_2$ having different ratios of initial $H_2$ $J$ states.   The main deviations between experiment and theory occur only at the most forward scattering angles where theory appears to overestimate the scattering intensity.   Both theory and experiments can be improved. For the theory part, the range of the PES sensitive to large impact parameters must be examined. For the experimental part, production of a pure $J$=0 population of $H_2$ or $D_2$ is needed, instead of a mixture including $J>0$ states. This will help directly clarify the differences in collision dynamics of $H_2O$-$H_2$($J$=0) versus $H_2O$-He. We have found strong differences in the DCSs between these two important collision systems, which play significant roles in the astrophysics applications.

From measurements of state-to-state relative integral cross sections, we note that the cross sections show an exponential decay with respect to transferred energy. This effect has been observed for $H_2O$-Ar inelastic collisions and attributed to the energy gap rule. The measured and calculated relative integral cross sections are in very good agreement. Both data indicate intramolecular alignment effects for $H_2O$ rotation during the collision. Collisions resulting in final rotational states of $H_2O$ with $Ka"=Kc"$ correspond to rotations around the $b$ axis of $H_2O$ and consequently have smaller cross sections. The dipolar transition propensity and energy gap rules dominates all other effects, when applicable. We also point out that the excellent agreement between theory and experiment for water-hydrogen (and water –helium) is not found for the relatively simpler water-Ar and water-Xe systems, as will be discussed in an upcoming publication. Clearly, the PESs for these larger rare gases are not correct in the regions probed by rotational inelastic scattering [16]. The lessons learned in improving PESs for such highly polarizable systems could have further implications for a better understanding of the collision dynamics of water.

**Acknowledgments**

The work in Nijmegen and in Grenoble was supported by the European FP6 program "Molecular

Universe". The Nijmegen group also acknowledges support by the Dutch National Science Foundation for Chemistry (NWO-CW ECHO 700.58.029), and by the NWO- Dutch Astrochemistry Network. The Institut de Planétologie et d'Astrophysique is a joint unit of C*NRS* and Université Joseph Fourier Grenoble. All calculations presented in this paper were performed at the Service Commun de Calcul Intensif de l'Observatoire de Grenoble (SCCI). This work was supported by the CNRS national programme "Physique et Chimie du Milieu Interstellaire", by the COST 'The Chemical Cosmos' program. LW acknowledges funding support from the NWO, as visiting scientist to Nijmegen.

**Figure Captions**

**FIG 1**. The rotational energy levels of ortho- and para-$H_2O$, plotted separately by the $Ka$ ladder. The shaded region shows the collision energy when using the $H_2$ beam with a 200K nozzle (collision energy 361 $cm^{-1}$). Ground state levels are indicated by a thick solid line while excited state levels populated by the collision and probed in this study are indicated by a thick dashed line. Principal rotational axes $a,b,c$ are labeled in the inset ball and stick model.

**FIG 2**. Schematic diagram of the crossed-beam velocity map imaging apparatus. A pulsed beam of $H_2O$ formed by seeding water vapor in Ar is skimmed and crossed by a second skimmed beam of pure normal- or para-$H_2$. The temperature of the secondary beam nozzle is fixed to either 200 or 320K. Rotationally excited $H_2O$ is state-selectively ionized by (2+1) REMPI via the $C$-state in the 248 nm region using a focused (20 cm lens) pulsed tunable dye laser beam that is frequency- doubled in a BBO crystal. The E field of the linear polarized laser beam lies perpendicular to the detector plane. The nascent $H_2O^+$ image is mass-selected by time-of-flight and projected onto a two dimensional (2-D) imaging detector then recorded by a CCD camera.

**FIG 3**. 3+1 REMPI spectrum of the ground vibronic state of $H_2$ via the $C$ (v=0) $\leftarrow X$ (v=0) transition. The black line represents the spectrum when using normal-$H_2$ while the red line shows the spectrum of our prepared para-$H_2$. Both spectra were taken with a 320 K nozzle.

**FIG 4**. Schematic illustration of the experimental geometry for studying inelastic collisions of $H_2O$ with $H_2$ ($J = 0, 1$, or $2$). End points of the vectors of the $H_2$ beam and the relative velocity are out of the range of the figure. The presented $H_2O^+$ image is a 2-D projection of the Newton sphere formed by ortho-$H_2O$ collisions with para-$H_2$, for the $1_{01} \rightarrow 2_{12}$ transition. 'CM' stands for the position of the center-of-mass and '0' is the crossing point of the two molecular beams ($H_2O$ and $H_2$ beams) and the laser beam. The center-of-mass vector connects the center-of-mass and lab frames. Forward scattering is defined as scattered $H_2O$ molecules moving along the direction of the $H_2O$ CM velocity.

**FIG 5.** Experimentally extracted (red curve) and calculated (black curve) state-to-state differential cross sections for $H_2O$ collisions with para-$H_2$ expanded in a 320K nozzle (collision energy 574 $cm^{-1}$). (a) ortho-$H_2O$ (b) para-$H_2O$. The experimental curves were scaled to match the theoretical curve at a deflection angle of 60°. Experimental uncertainties are estimated as <10% at all angles >5° for the stronger transitions and <20% for the weaker final states, e.g. the $3_{12}$, $4_{14}$, and $3_{13}$ states.

**FIG 6**. Comparison of experimental relative integral cross sections with quantum mechanical calculations for collisions of ortho- and para-$H_2O$ with para-$H_2$ (320K nozzle). The experimental relative cross sections are normalized to the calculated cross sections at the $1_{11}$ state. The experimental uncertainty is $\sim 20\%$.

**FIG 7**. Experimental state-to-state differential cross sections of ortho- and para-$H_2O$ colliding with normal-$D_2$, collision energy 430 $cm^{-1}$. All signals were normalized to set

each maximum intensity to unity.

**FIG 8.** Theoretical DCSs in $Å^2$/steradian for $H_2O+H_2$ scattering where four final rotational states for ortho- ($1_{01}\rightarrow1_{10},2_{12},2_{21}$) and para- ($0_{00}\rightarrow1_{11}$) $H_2O$ scattering are plotted for three different elastic events for $H_2$, namely $J=0\rightarrow J''=0$ (black curve), $2\rightarrow2$ (red curve), and $1\rightarrow1$ (green curve). The $J=0\rightarrow J''=0$ (black) curve is multiplied by 10 for the $1_{10}$, $1_{11}$, and $2_{12}$ final states for clarity.

**FIG 9**. Comparison of calculated and experimental DCSs for para- and normal-$H_2$ collisions with ortho- and para-$H_2O$ using the pulsed $H_2$ valve at 320K. Calculated DCSs: para-$H_2$ (black curve) and normal-$H_2$ (red curve); experimental DCSs: para-$H_2$ (green curve) and normal-$H_2$ (blue curve). Theory curves are in absolute units, while the experimental curves were scaled to match the theoretical curve at the deflection angle of 60°.

**FIG 10.** Experimental and calculated ratios of state-to-state relative cross section of $H_2O$ collision with normal- and para-$H_2$ using a 200K and 320K nozzle. The experimental results have 20% error bars and ~361 $cm^{-1}$ collision energy. The calculations consider rotational population for two different nozzle temperatures and collision energy of 575 $cm^{-1}$.

**FIG 11.** Comparisons of calculated integral cross sections for $H_2O$ inelastic collisions with $H_2$ (*J*, 0-0) and He at different collision energies. The calculation of $H_2O+H_2$ is from this work while calculations for $H_2O+He$ are from Patkowski *et al*.[37]

**Table Captions**

**Table 1.** Nozzle and rotational temperatures of $H_2$ molecular beam and the corresponding measured rotation populations.    Uncertainty is ~5% in estimations of the rotational temperature of $H_2$.

**Table 2.** Molecular beam conditions and collision energy.    Uncertainty is ~8% in speed ratio and velocity, and ~10% in collision energy.

**Table 3:** Calculated integral state-to-state cross section values for $H_2O + H_2$ inelastic transitions, at kinetic energy $E_{coll} = 575$ cm$^{-1}$ (320 K $H_2$ nozzle).    For each $H_2O$ state-to-state transition the cross section (in Å$^2$) for each $H_2$ state-to-state component is listed separately. The total $H_2O$ state-to-state cross section shown in Figure 6 are calculated for each $H_2O$ component by summing over the separate $H_2$ cross sections weighted by the $H_2$ state populations listed in

Table 1.

| Nozzle Temperature | Normal-H$_2$ | | Para-H$_2$ (90%) | |
|---|---|---|---|---|
| | 320K | 200K | 320K | 200K |
| Rotational Temperature Of H$_2$ | 220($\pm$10)K | 170($\pm$10)K | 220($\pm$10)K | 170($\pm$10)K |
| $J$=0 | 17($\pm$0.5)% | 20($\pm$0.6)% | 61($\pm$2.2)% | 73($\pm$2.5)% |
| $J$=1 | 72($\pm$0.5)% | 74($\pm$0.3)% | 10($\pm$0.1)% | 10($\pm$0.1)% |
| $J$=2 | 8($\pm$0.5)% | 5($\pm$0.7)% | 29($\pm$2.1)% | 17($\pm$2.5)% |
| Higher $J$ | 3($\pm$0.5)% | 1($\pm$0.3)% | <1% | <1% |

Table 2.

| Nozzle conditions | Secondary beam | | | Primary beam |
|---|---|---|---|---|
| | He | $D_2$ | $H_2$ | $H_2O$ (seeded in Ar) |
| Nozzle pressure (bar) | 1 | 1 | 1 | 1 |
| Nozzle Temperature (K) | 320 | 320 | 320 <br> 200 | 320 |
| Peak velocity (m/s) | 1660 | 1660 | 2700 <br> 2100 | 625 |
| Speed ratio | 8 | 8 | 6 | 10 |
| Collision energy ($cm^{-1}$) | 430 | 430 | 574 <br> 361 | - |

Table 3

| H$_2$O transition $J_{Ka Kc} \rightarrow J''_{Ka'' Kc''}$ Energy level (cm$^{-1}$) | H$_2$ transition $J \rightarrow J''$ | Cross section (Å$^2$) | H$_2$O transition $J_{Ka Kc} \rightarrow J''_{Ka'' Kc''}$ Energy level (cm$^{-1}$) | H$_2$ transition $J \rightarrow J''$ | Cross section (Å$^2$) |
|---|---|---|---|---|---|
| *ortho* **1$_{01}$ → 1$_{10}$** | 0 → 0 | 4.22 | *Para* **0$_{00}$ → 1$_{11}$** | 0 → 0 | 8.09 |
| 23.799 | 2 → 2 | 7.68 | 0 | 2 → 2 | 15.10 |
| 42.402 | 1 → 1 | 8.26 | 37.158 | 1 → 1 | 15.71 |
|  | 0 → 2 | 0.074 |  | 0 → 2 | 0.061 |
|  | 2 → 0 | 0.167 |  | 2 → 0 | 0.307 |
| **→ 2$_{12}$** | 0 → 0 | 4.49 | **→ 2$_{02}$** | 0 → 0 | 1.39 |
|  | 2 → 2 | 8.48 |  | 2 → 2 | 5.42 |
| 79.513 | 1 → 1 | 9.12 | 70.133 | 1 → 1 | 5.70 |
|  | 0 → 2 | 0.062 |  | 0 → 2 | 0 |
|  | 2 → 0 | 0.269 |  | 2 → 0 | 0.631 |
| **→ 2$_{21}$** | 0 → 0 | 0.562 | **→ 2$_{11}$** | 0 → 0 | 0.018 |
|  | 2 → 2 | 2.22 |  | 2 → 2 | 1.67 |
| 135.322 | 1 → 1 | 2.71 | 95.245 | 1 → 1 | 1.51 |
|  | 0 → 2 | 0.011 |  | 0 → 2 | 0 |
|  | 2 → 0 | 0.199 |  | 2 → 0 | 0.202 |
| **→ 3$_{21}$** | 0 → 0 | 0.265 | **→ 2$_{20}$** | 0 → 0 | 0.784 |
| 212.628 | 2 → 2 | 1.19 |  | 2 → 2 | 3.34 |
|  | 1 → 1 | 1.36 | 136.588 | 1 → 1 | 4.09 |
|  | 0 → 2 | 0.002 |  | 0 → 2 | 0.006 |
|  | 2 → 0 | 0.501 |  | 2 → 0 | 0.509 |